\documentclass[11pt]{article}

\usepackage[a4paper, margin=2cm]{geometry}
\usepackage{soul}
\usepackage{booktabs}
\usepackage{multirow}
\usepackage{xspace}
\usepackage{amssymb}
\usepackage{authblk}
\usepackage{graphicx}

\newcommand{\bps}{BPS-BBP\xspace}
\newcommand{\hp}{HP-DH\xspace}
\newcommand{\vbb}{SCVBB\xspace}

\begin{document}

\title{Consistent crust-core interpolation and its effect on
  non-radial neutron star oscillations}

\author[1,2]{Mart\'in O. Canull\'an-Pascual \footnote{canullanmartin@fcaglp.unlp.edu.ar}}
\author[1,2]{Mauro Mariani}
\author[1,2]{Ignacio F. Ranea-Sandoval}
\author[1,2]{Milva G. Orsaria}
\author[3,4]{Fridolin Weber}

\affil[1]{Grupo de Astrof\'isica de Remanentes Compactos, Facultad de
  Ciencias Astron\'omicas y Geof\'isicas, Universidad Nacional de La
  Plata, Paseo del Bosque S/N, La Plata (1900), Argentina}

\affil[2]{CONICET, Godoy Cruz 2290, Buenos Aires (1425), Argentina}

\affil[3]{Department of Physics, San Diego State University (SDSU),
  San Diego, CA, United States of America}

\affil[4]{Department of Physics, University of California at San
  Diego (UCSD), La Jolla, CA, United States of America}

\date{}
\maketitle

\abstract{To model the structure of neutron stars (NSs) theoretically,
  it is common to consider layers with different density regimes.
  Matching the equation of state (EoS) for the crust and core and
  obtaining a suitable description of these extreme conditions are
  crucial for understanding the properties of these compact objects.
  In this work, we construct ten different NS EoSs incorporating three
  distinct crust models, which are connected to the core using a
  thermodynamically and causally consistent formalism. For cold NSs,
  we propose a linear relationship between pressure and energy density
  in a narrow region between the crust and core, effectively
  establishing an interpolation function in the pressure-baryonic
  chemical potential plane. We then compare this EoS matching method
  with the classical approach, which neglects causal and thermodynamic
  consistency.  We solve the Tolman–Oppenheimer–Volkoff equation to
  obtain the mass-radius relationship and compare our results with
  observational constraints on NSs. Furthermore, we investigate the
  influence of the new matching formalism on non-radial oscillation
  frequencies and damping times. Our findings suggest that the method
  used to \textit{glue} the crust and core EoS impacts NS observables,
  such as the radius, oscillation frequencies, and damping times of
  non-radial modes, which may be crucial for interpreting future
  gravitational wave observations from neutron star mergers or
  isolated pulsars. The effects are particularly noticeable for
  low-mass NSs, regardless of the specific EoS model chosen. In
  particular, we find that the $p_1$ oscillation mode exhibits
  significant differences in frequencies among alternative matching
  methods, whereas the fundamental $f$-mode remains unaffected by
  changes in crust models or interpolation schemes.}

\begin{center}
\end{center}

\section{Introduction}

Neutron stars (NSs) are among the densest objects in the Universe,
reaching central densities several times above the nuclear saturation
density. Despite the fact that most of the theoretical uncertainties
are related to the behavior of matter at densities above this value,
there are still some open questions connected to the crust of these
intriguing objects.

Only a small percentage of the mass of NSs is contained within the crust.
Despite this, it is proven crucial for many astrophysical
phenomena related to these compact objects. For example, it is key in
the determination of the radius and, for that reason, objects
constructed with different crust models might present distinctive
merging properties.

Moreover, with the theoretical and technological advancements, the
last decade revolutionized our capabilities of studying these
mysterious and intriguing compact objects. Direct detection of merging
compact stars
\cite{Abbott:2017mmo,Abbott:2017oog,Abbott:2018exr,Abbott:2020goo},
estimates of masses and radii of isolated NSs with the NICER telescope
\cite{Miller:2019pjm,Riley2019anv,Miller:2021tro,Riley:2021anv}, and
 observations of $2\,M_\odot$ NSs
\cite{Antoniadis:2013amp,Demorest:2010ats,Arzoumanian:2018tny,Fonseca:2021rfa}
has allowed us to constrain some aspects of the equation of state
(EoS) of dense matter \cite{Fattoyev:PRC.2020}.
\begin{figure}
  \includegraphics[width=.8\linewidth]{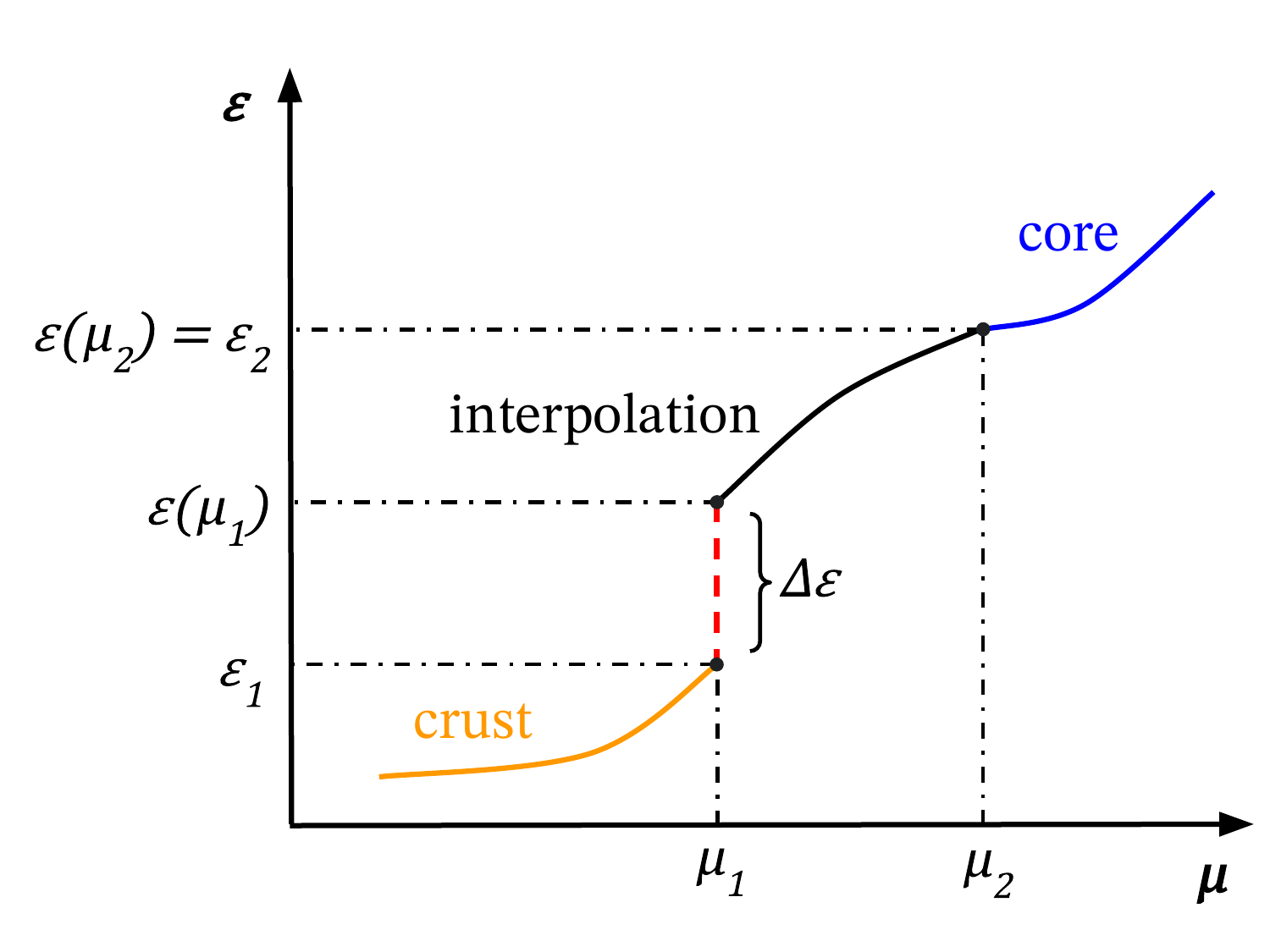}
\caption{Illustrative representation of the energy density jump at the crust-interpolation region. Continuity of the energy density at the interpolation-core region is ensured by selecting the minimum value of the speed of sound, $c_s^2$, which guarantees causal and thermodynamic consistency. The number density interpolation, $n(\mu)$, exhibits analogous behavior, also presenting a single discontinuity at the crust-interpolation region.}
    \label{fig:esquema}
\end{figure}

Neutron stars can typically be described as consisting of several
layers with distinct density regimes, with the simplified model of the
outer and inner crusts and cores providing a representative framework
for understanding their structure.
 Each of these layers has unique properties that can influence
various observational features of NSs, such as their interaction with
magnetic fields, mass-radius relationship, and non-radial oscillations
\cite{Haensel:2007nst}. From the crust to the core, NSs pass through
different density regimes that must be described mainly in terms of
strong interactions. However, due to the complexity of quantum
chromodynamics, there is no single EoS that can describe all the
layers mentioned above. The EoS for the different layers of a NS is
typically matched by using a low-density crust EoS, relatively well
constrained, and a high-density core EoS, still unknown.

In this work, we present a consistent method to correctly match the
EoSs of the crust and core of NSs. We explore how these techniques
impact on the non-radial oscillation modes of NSs, which are
responsible for generating gravitational waves. These waves, which
travel through the universe and reach our detectors, can provide
valuable information about the properties of compact
objects. Specifically, the frequencies and damping times of these
oscillation modes depend on the stellar structure and composition of
NSs.

This work is organized in the following manner. In Section~\ref{sec.2}
we present particularities of the equation of state, focusing
particularly in the {\it{gluing}} formalisms. In Section~\ref{sec.3}
we present macroscopic results obtained using the different
interpolation methods and those associated with non-radial modes,
specifically for the $p_1$-mode. This mode is
vital for interpreting gravitational wave signatures
from neutron star mergers or isolated oscillating neutron stars as they
 provide a direct probe into the star’s interior structure and
the properties of its EoS.
Section \ref{sec.4} is devoted to
draw the main conclusions and astrophysical implications of our work.

\section{Crust-Core EoS interpolation} \label{sec.2}

\begin{table}[t]
\centering
\begin{tabular}{ccc}
\toprule Model & $n_1$ & $n_2$ \\ \midrule \bps-a & $n_0/2$ &
\multirow{2}{*}{$\frac{3}{4}n_0$} \\ \bps-b & $n_0/10$ & \\ \midrule
\hp-a & $n_0/10$ & \multirow{2}{*}{$\frac{2}{3}n_0$} \\ \hp-b &
$0.01$~fm$^{-3}$ & \\ \midrule \vbb-a & $0.01$~fm$^{-3}$ &
\multirow{3}{*}{$\frac{1}{2}n_0$} \\ \vbb-b & $n_c$ & \\ \vbb-c &
$n_0/10$ & \\ \bottomrule
\end{tabular}
\caption{Constructed crust models with their respective selected $n_1$ and $n_2$ parameters. The nuclear saturation density is $n_0 = 0.16~\mathrm{fm}^{-3}$, and $n_c$ represents the crossing density between the crust and hadron EoSs, which in this case is $n_c = 0.135~\mathrm{fm}^{-3}$.}
\label{table:EoS_densities}
\end{table}

\begin{figure*}[t]
\centering
\includegraphics[width=\linewidth, height=0.25 \textheight]{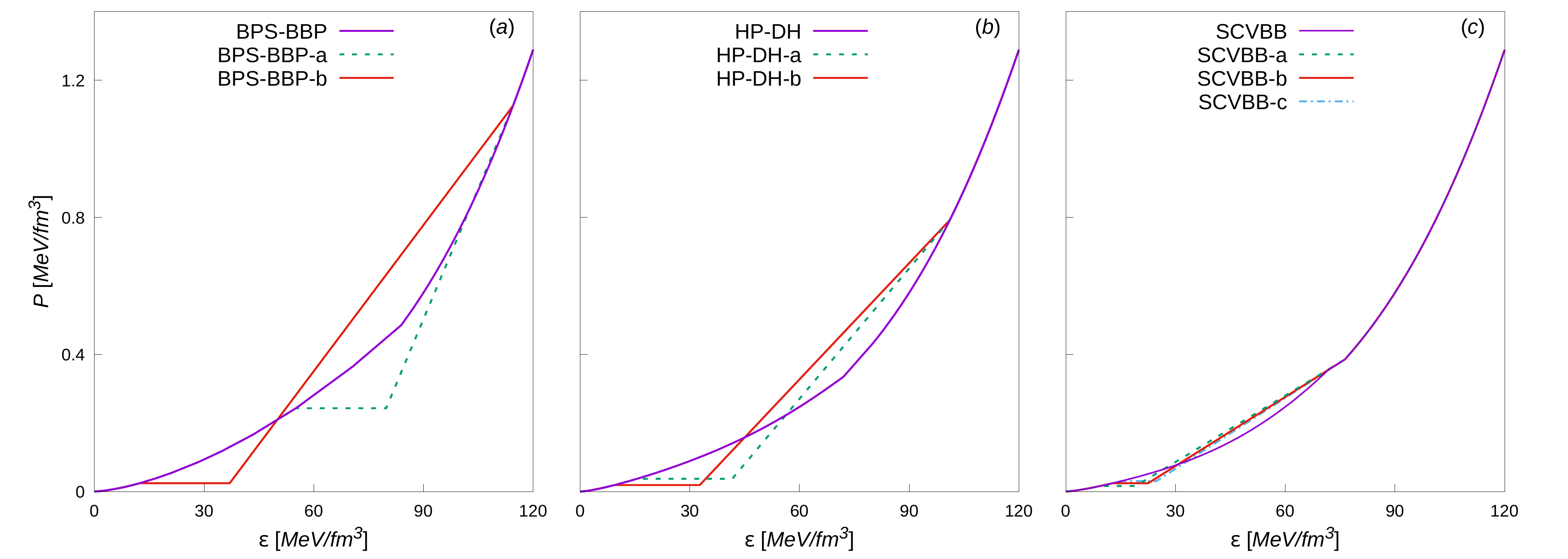}
\caption{Details of the \bps, panel $(a)$; \hp, panel $(b)$; and \vbb, panel $(c)$, EoSs for crust densities. Results are presented for the different interpolations detailed in Table~\ref{table:EoS_densities} as well as for the traditional non-interpolated construction (see text for details). In panels $(a)$ and $(b)$, the thermodynamically consistent EoSs exhibit an energy density gap, $\Delta \epsilon$, of some tenths of MeV/fm$^3$ after the crust ends, in contrast to the non-consistent constructions. For the \vbb~EoS sets in panel $(c)$, the obtained $\Delta \epsilon$ values are smaller than those for the other crust models, resulting in smaller differences with the non-interpolated version. Among all the constructed EoSs, only the \bps-a parametrization presents a slightly higher transition pressure.}
\label{fig:eos}
\end{figure*}

In this section, we present the thermodynamical and causal consistent
formalism for gluing the NSs crust and core EoSs. Following the work
of \cite{fortin2016nsr}, we use an interpolation scheme for the
pressure as a function of the chemical potential in the transition
region between the crust and core.  To describe the matter in the core
of NSs populated by nucleons, hyperons, and delta resonances, we used
a specific parametrization of the density dependent relativistic mean
field theory, the SW4L parametrization, along with electrons and muons
for the leptonic contribution, details of this parametrization
  can be found in \cite{Malfatti:2020dba, Celi:2024doh}. For the crust,
we used the \bps \cite{baym1971tgs,baym1971nsm}, the \vbb
\cite{sharma2015ueo} and the \hp \cite{Haensel1994enm,
  douchin2001aue} models acquired from the literature.   The chosen EoS
models represent a broad spectrum of theoretical approaches to
modeling neutron star crust and core properties. For instance, the
\bps model emphasizes semi-empirical nuclear mass data, while the
\vbb incorporates modern self-consistent treatments of nuclear pasta
structures, capturing key theoretical uncertainties in dense matter
physics.

The classical, most applied way of combining EoSs of the crust and the
core is to simply concatenate the selected crust EoS table with the
hadronic one, ensuring that the thermodynamic quantities increase
monotonically. In a more consistent approach, \cite{fortin2016nsr}
  assume that in the crust-core transition region the pressure-energy
density relation can be approximated by a linear interpolating
function. Its slope{, the square of the speed of sound $c_s^2$ in the
  interpolating region,} is the only free parameter of the model, {and
  it has been proven to have} a minimum value for which the
interpolation ensures thermodynamic and causal consistency {(see
  details below)}.

Thermodynamic and causal consistency require that the following general conditions on the derivatives be satisfied in the interpolation region:
\begin{itemize}
\item {The baryonic density, $n$, must be an increasing function of pressure, P, meaning that P($\mu$), with $\mu$ the chemical potential, is both increasing and convex;}
\item {The condition $(\partial P/\partial \varepsilon)^{1/2}$ = $c_s/c < 1$, holds with the energy density given by {the Euler relationship,} \mbox{$\varepsilon(\mu) = n(\mu)\mu - P(\mu)$}.}
\end{itemize}

\begin{figure*}[t]
    \centering
    \includegraphics[width=\linewidth, height=0.25 \textheight]{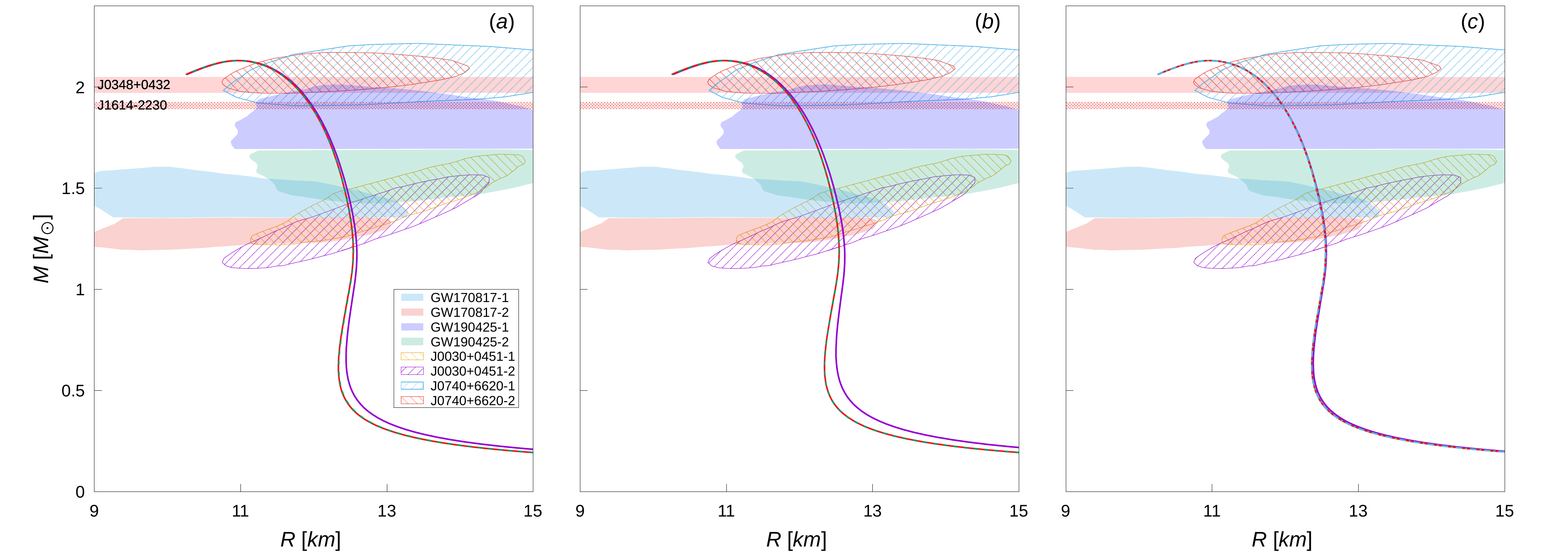}
\caption{Mass-radius relationships for the \bps, panel $(a)$; \hp, panel $(b)$; and \vbb, panel $(c)$, EoSs. The same color coding for curves as in Fig.~\ref{fig:eos} is used. Modern astronomical constraints are also shown as colored regions. In all cases, within each panel, the main differences appear in the radii of low-mass NSs between the interpolated models and the non-interpolated one. Additionally, within each panel, the differences between the various interpolated parametrizations are negligible. Notably, for the \vbb EoS sets, as discussed in the $P$-$\epsilon$ relationship, the very small differences between the interpolated and non-interpolated methods correspond to almost unnoticeable differences in the $M$-$R$ plane.}
    \label{fig:mr}
\end{figure*}

This technique introduces a potential energy density gap, i.e., a
first-order phase transition, in the transition between the crust and
the interpolation sector, as well as between the interpolation sector
and the core.\footnote{{It is interesting to
  mention that, if one makes a linear interpolation so that it matches
  both start and ending points of the EoSs, you will never reach a
  continuity in the speed of sound, unless it has the same value at
  both points.}}; within the method by \cite{fortin2016nsr}, this
potential features are determined once selected the slope
parameter.

Since the chemical potential, $\mu$, is continuous across all sectors
(including potential density gaps), the pressure, $P$, number density,
$n$, and energy density, $\varepsilon$, are modeled as functions of
$\mu$.  As $\mu_1$, all subscripts $1$ correspond to the last point of
the crust region, and as $\mu_2$, all subscripts $2$ correspond to the
first point of the core region.

If one select for the slope parameter the previously mentioned minimum value, the baryon number and energy densities are continuous at the transition between the interpolating region and the core,
\begin{equation}
    n_2=n(\mu_2), \quad \varepsilon_2=n(\mu_2) \, ,   
\end{equation}
and there appear only one density gap on the EoS at the transition between the crust and the interpolation sector. In this region, at the crust ending, we obtain: 
\begin{equation}
n_1\neq n(\mu_1), \quad  \varepsilon_1\neq \varepsilon(\mu_1) \, .
\end{equation}
A semiaquatic illustration is shown in Fig.~\ref{fig:esquema}

In this work, we focus solely on this specific case of the minimum consistent value of $c_s^2$,  which, as previously mentioned, results in a single jump discontinuity on the energy densities, $\varepsilon$, and the baryon density, $n$, where the interpolation begins and the crust ends.

Furthermore, a thermodynamic requirement is that the pressure, $P$,
must always remain continuous respect to the chemical potential, which serves as
the independent variable in the interpolation region:
\begin{equation}
P_1=P(\mu_1), \qquad P_2=P(\mu_2) \, .
\end{equation}
Thus, the corresponding interpolation function is given by:
\begin{equation}
    P(\mu)=P_1+\Delta P \frac{(\mu^{b}-\mu_1^{b})}{(\mu_2^{b}-\mu_1^{b})} \, ,
    \label{eq1}
\end{equation}
where $\Delta P=P_2-P_1$ and $b=1+1/c_s^{2}$. In the interpolation
region, $c_s^{2}$ is held constant, as required by the formalism.

Imposing monotonicity of $\varepsilon(\mu)$  over the entire range of the interpolation region, there appear two non-linear equations:
\begin{equation} \label{Eqb1}
    \frac{b}{q-1}-\frac{n_1\mu_1}{\Delta P}=0 \, ,
\end{equation}
\begin{equation} \label{Eqb2}
    \frac{bq}{q-1}-\frac{n_2\mu_2}{\Delta P}=0 \, ,
\end{equation}
with $q=(\mu_2/\mu_1)^b$. Solving them separately, we obtain two solutions for $b$, $b_1$ (from Eq. (\ref{Eqb1})) and $b_2$ (from Eq. (\ref{Eqb2})). If $b_1 > b_2 $ then, it is the scenario where there is a single density jump at $n_1$. In contrast, if $b_1< b_2$, the density jump occurs only at $n_2$. Then, when imposing monotonicity of $\varepsilon(\mu)$, an upper bound appears on $b$, $b_{max}=\mathrm{min}(b_1,b_2)$, and consequently the {previously introduced} minimum possible value for the slope appears, ${c^2_{min}}=\mathrm{max}({c^2_1},{c^2_2})$, being $c^2_1$ and $c^2_2$ the values associated with $b_1$ and $b_2$, respectively. Finally{, in the general case,} selecting a value above ${c^2_{min}}$ will give two density jumps, one at $n_1$ and the other at $n_2$.

\begin{figure*}
    \centering
    \includegraphics[width=\linewidth, height=0.25 \textheight]{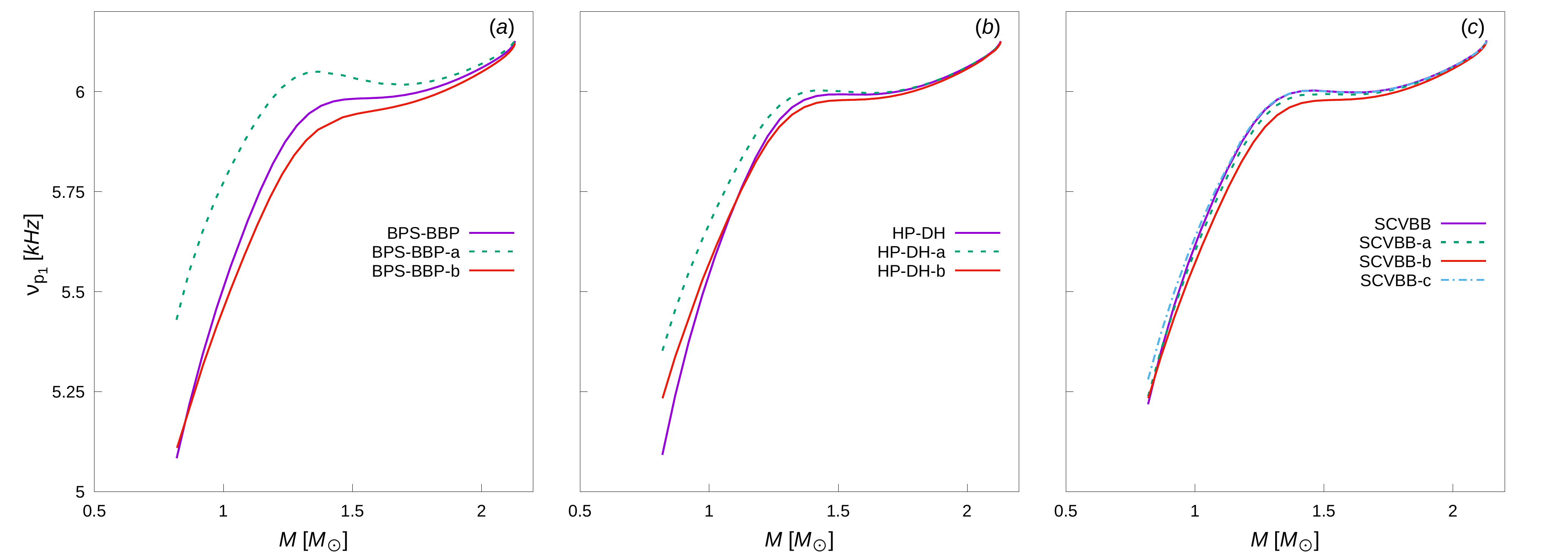}
    \caption{Frequency-mass relationships for the $p_1$-mode for the \bps~(panel $(a)$), \hp~(panel $(b)$), and \vbb~(panel $(c)$) EoSs. In all panels, noticeable differences among the sets are evident. Specifically, for the \bps~EoSs in panel $(a)$, a difference of approximately $300$~Hz can be observed for a $1\,M_\odot$ NS between the different interpolated parametrizations, with the non-interpolated case lying in between. Additionally, smaller differences are apparent even in the high-mass regions. Similar qualitative behavior is seen in panels $(b)$ and $(c)$ for the \hp and \vbb EoSs, respectively, although the \vbb~EoSs exhibit the smallest differences among the sets. A global increasing trend in frequency is observed with increasing stellar mass, although the curves exhibit a local minimum in certain regions.}
    \label{fig:nu}
\end{figure*}

Therefore, Eq. (\ref{eq1}), allows to compute $P$ and $\varepsilon$
  from the proposed linear relation:
\begin{equation}
P=c_s^{2}(\varepsilon-\tilde{\varepsilon}) \, , 
\end{equation}
where 
\begin{equation}
\tilde{\varepsilon}=b\frac{P_2-qP_1}{q-1}\, ,
\end{equation}
obtained from continuity of $\mu$, and $n$ from the Euler relation,
\begin{equation}
n=\frac{(P+\varepsilon)}{\mu} \, ,
\end{equation}
which is valid throughout the entire interpolation region guaranteeing thermodynamic consistency. Causal consistency is achieved {imposing} the speed of sound, $c_s^2$, to be less than speed of light in vacuum. The baryon density, $n$, can be obtained from the derivative of $P(\mu)$ given by Eq. (\ref{eq1}):
\begin{equation}
    \frac{dP}{d\mu}=\frac{b\Delta P}{(\mu_2^{b}-\mu_1^{b})}\mu^{1/c_s^{2}}=\frac{b\Delta P}{(q-1)\mu_1}{\Big(\frac{\mu}{\mu_1}\Big)}^{1/c_s^{2}} \, .
\end{equation}

Within this thermodynamic and causal consistent interpolation technique, we construct seven different parametrizations, two using the \bps crust, two using the \hp crust, and three using the \vbb crust. The specific values chosen for $n_1$ and $n_2$ in this EoS are presented in Table~\ref{table:EoS_densities}. These choices were selected following the possibilities presented by \cite{fortin2016nsr}. We also construct three additional EoSs, one for each of the crust models mentioned, using the traditional concatenation technique.

\section{Results} \label{sec.3}

Given the mentioned sets of selected EoSs, we integrated the TOV equation, which describe the hydrostatic equilibrium of spherically symmetric stars in general relativity; from this, we obtained our theoretical mass-radius curves. In order to calculate non-radial oscillations modes, we used the formalism developed by \cite{lindblom1983tqo} and \cite{Detweiler1985otn} where the linearized Einstein equations (together with proper boundary conditions) are numerically solved. In this manner, the frequencies and damping times of non-radial oscillation modes are obtained. The results obtained after all of these calculations for our sets are presented below.

\begin{figure*}
    \centering
    \includegraphics[width=\linewidth, height=0.25 \textheight]{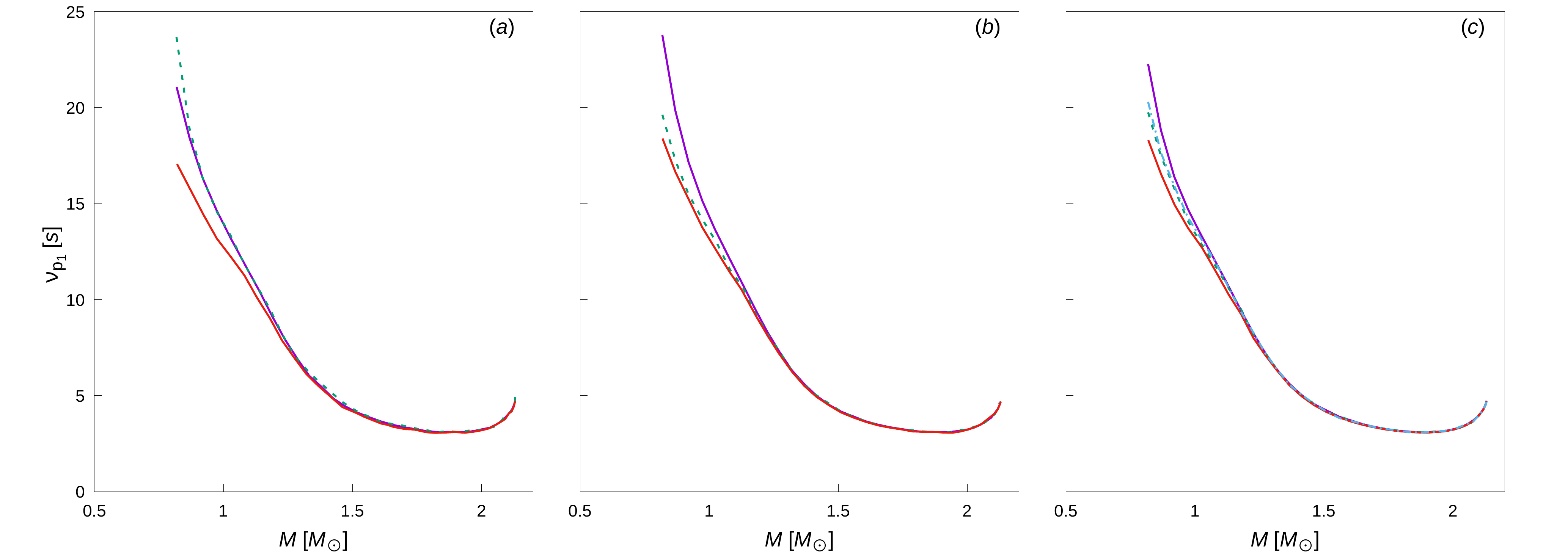}
\caption{Damping time-mass relationships for the $p_1$-mode for the \bps~(panel $(a)$), \hp~(panel $(b)$), and \vbb~(panel $(c)$) EoSs, with the same color coding as in Fig.~\ref{fig:nu}. Compared to the $\nu$ results, the differences among the sets for $\tau$ are smaller and, for high masses, are almost indistinguishable. All panels display the same global morphology for the damping time curves, with differences among the sets appearing only for low-mass stars. Unlike previous figures, the \vbb results for $\tau$ do not exhibit the smallest differences among the sets and are not significantly distinct from the \bps and \hp results.}
    \label{fig:tau}
\end{figure*}

In Fig.~\ref{fig:eos} we present the crust region of the pressure-energy density relationship for these ten EoSs. There it can be seen how the thermodynamical and causal{{ly consistent}} method selected introduce an energy density gap (sign of a first order phase transition), in contrast to the non-consistent concatenate method, which produces a monotonous continuous crust-core transition.
By design, the \bps and \hp parametrizations (panels $(a)$ and $(b)$,
respectively) produce greater differences with their corresponding
non-interpolated EoSs than the \vbb sets (panel $(c)$).
 The main reason for this is that the interpolation region of SCVBB sets is smaller in terms of the baryon density interval.

In Fig.~\ref{fig:mr} we present the mass-radius relationships for the ten {corresponding} EoSs. We also present in this figure the current astronomical constraints; they represent the detected NSs, coming from $2~M_\odot$ pulsars, gravitational waves NS merger events and X-ray NICER detections, and sampled by their radius and masses, with the respective measurement errors. We can appreciate, {particularly clearly in panels $(a)$ and $(b)$ of} Fig.~\ref{fig:mr}, small differences between the classical and the new formalism for low-mass NSs. {It is worth mentioning that within the classical approach to match the crust and the core EoSs, the choice of the \textit{gluing} baryonic density produces important differences in low-mass NSs.} The maximum mass reached for a stable star in each curve of any of Fig.~\ref{fig:mr} is not affected by the chosen crust-core matching {formalism}. However, this is not the case for the radius and crust thickness of stars with $M \lesssim 1~M_\odot$, which correspond to typical NS mass configurations. This is particularly noticeable for \bps and \hp sets, panels $(a)$ and $(b)$ respectively, while \vbb sets, panel $(c)$, present smaller differences due to the similarities among \vbb EoSs (see panel~$(c)$ of Fig.~\ref{fig:eos}).

On the other hand, we calculated the frequencies, $\nu$, and damping times, $\tau$, of non-radial modes for all our sets. We found that, the differences among our EoS sets, arising from the different crust-core interpolations, does not have a significative impact on the fundamental $f$-mode, but it does have an impact on first pressure $p_1$-mode; thus, we only present results for the $p_1$-mode. In Figs.~\ref{fig:nu} and \ref{fig:tau} we present the $\nu$-$M$ and $\tau$-$M$ relationships for this mode; as previously done, we show results for the different crust models, \bps, \hp, and \vbb, in three contiguous panels. For the $\nu$ results in Fig.~\ref{fig:nu}, we can see that there appear noticeable differences, even for higher masses; in this sense, the frequency of the $p_1$-mode appears to be the most sensible quantity to the crust-core interpolation method. As an extreme case, there appears a difference of $300$~Hz for the \bps model, between the \bps-a and \bps-b EoSs. The other panels, presenting \hp and \vbb EoSs, show smaller differences, but still non-negligible.
In all panels, as the mass approaches the maximum value, the  $\nu$  values of the different sets converge to the same result.
 Regarding results in Fig.~\ref{fig:tau}, all panels show a similar behavior, where damping times for $p_1$-mode are approximately the same for compact objects with high masses, and only small differences appear for low-mass configurations. {The damping times are not as sensitive to the crust-core transition as the frequencies are.}

\section{Conclusions and discussion} \label{sec.4}

We studied the astrophysical effects of the \textit{gluing} formalism
between the crust and core EoSs in a thermodynamic and causal
consistent way, focusing on the impact on NSs non-radial oscillation
modes. We considered three different EoS for the crust: the classical
BPS-BBP, where the nuclear masses for the outer crust are derived from
a semi-empirical mass table, and the inner crust EoS is based on
Brueckner and variational calculations for pure neutron matter; the
HP-DH, which is based on experimental nuclear data and Skyrme Lyon
(SLy) effective nucleon-nucleon interactions; and SCVBB, an improved
BPS EoS including the Brueckner-Hartree-Fock approach for nucleon
interactions, with the inner crust modeled using Thomas-Fermi
calculations, including nuclear pasta structures
self-consistently. For the outer and inner cores of NSs, we used the
SW4L hadronic EoS.

We construct seven different parametrizations choosing different
crustal thicknesses, ensuring a consistent thermodynamic and causal
match with the SW4L core EoS. For comparison, we also apply the
traditional concatenation technique and contrast it with the EoS
obtained from the seven parametrizations.

The consistent interpolation method introduces a gap in the energy
density, indicative of a first-order phase transition, in contrast to
the usual non-consistent concatenation technique, which results in a
monotonous, continuous crust-core transition.

Our main results indicate that, regardless of the specific crust
models we employ, the formalism used to construct the complete EoS has
a significant impact on the $p_1$-mode frequencies for low-mass NSs,
while the $f$-mode remains unaffected. We hypothesize that this
difference can be attributed to the distinct characteristics of the
$p_1$-mode eigenfunctions, which are highly concentrated near the
star's surface, particularly in the region where the crust and core
transition occurs. This proximity to the core-crust boundary likely
amplifies the sensitivity of the $p_1$-mode to the details of the
crust's EoS.

The significant variations in the $p_1$-mode frequency for low-mass
neutron stars between interpolated and non-interpolated methods
suggest that these modes are highly sensitive to the crust-core
transition region. This sensitivity could provide a unique
observational diagnostic for distinguishing among different crust
models and interpolation schemes.

In this sense, the frequency of the $p_1$-mode stands out as the most
sensitive probe of the crust-core transition region properties.
Unlike other quantities, such as the damping time of the same mode or
the frequencies of other relevant modes, like the fundamental $f$-mode,
the $p_1$-mode uniquely reflects changes in the crust’s EoS.
 Additionally,
macroscopic observables such as the mass or radius of the neutron star
are far less affected by the details of the crust-core transition. Our
findings align with those of \cite{Kunjipurayil:2022iot}, which
similarly highlight the localization of the $p_1$-mode eigenfunctions
near the star’s surface. This localization makes the $p_1$-mode
particularly sensitive to variations in the crust-core boundary,
amplifying the impact of differences in the crust-core construction.

In the future, simultaneous measurements of NS mass and radius with
high precision, potentially within a few percent, may be achieved
through X-ray emission analyses, such as those performed by NICER.
 Additionally, with the advent of next-generation
gravitational wave detectors, it is anticipated that gravitational
waves emitted by isolated NSs will be detected, enabling the
estimation of their non-radial oscillation modes. Consequently, it is
crucial to analyze and thoroughly address potential observables of
NSs, with particular focus on modeling and exploring the uncertainties
and intricacies of the EoS of the dense matter composing these compact
objects.

Prospective studies should explore the relevance of $p_1$-mode sensitivity for
constraining neutron star interiors, particularly in the context of
future gravitational wave and X-ray observations. With advancements in
gravitational wave detectors and X-ray missions like NICER, the
ability to simultaneously measure mass, radius, and oscillation
frequencies will provide unprecedented insights into the equation of
state of dense matter. The $p_1$-mode, being highly sensitive to the
crust-core transition, stands out as a unique probe for these critical
properties.

Additionally, extending this research to incorporate
temperature-dependent effects could refine our understanding of
thermal influences on crust-core interpolation and neutron star
observables. Investigating alternative crust models and their impact
on interpolation schemes would also help in capturing the diversity of
possible neutron star compositions. These efforts will further advance
the field of multi-messenger astrophysics and improve our
understanding of dense matter physics.

\section*{Acknowledgments}

MOC-P is a doctoral fellow at CONICET (Argentina). MOC-P, MM, IFR-S,
and MGO acknowledge UNLP and CONICET (Argentina) for financial support
under grants G187 and PIP 0169. IFR-S is also partially supported by
PIBAA 0724 from CONICET.

\end{document}